\documentclass[a4paper,11pt]{article}
\usepackage{amssymb,amsmath,bm}  % for math
\usepackage{graphics,graphicx}   % for figures
\usepackage{array,booktabs}      % for tables
\usepackage{authblk}  % for footnote style author/affiliation

\usepackage{cite}

\usepackage[margin=2.5cm]{geometry}

\usepackage[linktocpage,
colorlinks = true,
linkcolor = blue,
urlcolor  = blue,
citecolor = red,
anchorcolor = blue]{hyperref}

\usepackage{epsfig}

\DeclareUnicodeCharacter{2212}{-}
\DeclareUnicodeCharacter{202F}{\,}

\numberwithin{equation}{section}

\title{\textbf{Thermal Leptophilic Light Vector Dark Matter with Spinor Mediator and Muon (g-2) Anomaly}}
\author[1]{Seyed Yaser Ayazi\thanks{syaser.ayazi@semnan.ac.ir}}
\author[2]{Ahmad Mohamadnejad\thanks{mohamadnejad.a@lu.ac.ir}}
\affil[1]{Physics Department, Semnan University, P.O. Box. 35131-19111, Semnan, Iran}
\affil[2]{Department of Physics, Lorestan University, khorramabad, Iran}

\date{\today}

\begin{document}

\baselineskip 0.6 cm

\maketitle

\begin{abstract}
Inspired by the recently new measurement of $(g-2)_{\mu}$ at FermiLab and reported upper bound  for electron-dark matter (DM) recoil  by the XENON1T collaboration, we revisited phenomenology of a light MeV scale vector dark  matter in a leptophilic extension of standard model while a new spinor field plays the role of mediator. A viable parameter space is considered to discuss the possibility of light dark matter relic density as well as anomalous magnetic moment of the muon. We study DM-electron direct detection and cosmological bounds on the parameters space of the model. It is shown that although new bound of $(g-2)_{\mu}$ anomaly greatly confines the parametric space of the model, the thermal light dark matter can exist for $\rm M_{DM}\sim 10^{-1}-10^{1}~\rm GeV$.

\end{abstract}

%\newpage

%\tableofcontents

\section{Introduction} \label{sec1}
As it is well-known, the existence of DM on many scales in the universe has been widely established, but its nature remains an unsolved problem. Traditional DM candidates, in many beyond  standard model (BSM) scenarios are weakly interacting massive particle (WIMP) with mass close to the electroweak scale and weak couplings with Standard Model (SM) particles. However the lack of any signal in direct, indirect detection experiments and search at the LHC, has opened the door to wide spreading DM models possibility  covering small value ranges in  DM masses and interaction coupling strengths. 

Furthermore, physics of BSM may leave affect in low-energy precise measurements via possible deviations between the SM and experimental evidences. Some of these deviations include the LHCb \cite{LHCb:2021trn} and the muon anomalous magnetic  moment (AMM) $a_{\mu}=(g_{\mu}-2)/2$  which was recently reported by FermiLab experiment ($3.3~\sigma$)\cite{Muong-2:2021ojo} challenging lepton  universality. Combining this result for AMM with the earlier measurement
at the Brookhaven National Laboratory, this now amounts to a $4.2~\sigma$ deviation from the SM prediction \cite{Muong-2:2006rrc}. This result leaves an open window to new particles which can contribute to muon $g-2$ through direct coupling to the muon. Many models related to DM candidates and $(g-2)_{\mu}$ anomaly have been proposed, including axion-like particle (ALP) explanation \cite{deNiverville:2018hrc,Chiang:2018bnu,Keung:2021rps,Bhattacharya:2021shk}, SUSY-DM explanation \cite{Shafi:2021jcg,Forster:2021vyz,Frank:2021nkq,Kim:2021suj}, two Higgs doublet models (2HDM) plus singlet explanation \cite{Arcadi:2021yyr}, and others \cite{Cacciapaglia:2021gff,Jia:2021mwk,Singirala:2021gok,Qi:2021rhh,NA64:2021acr,Saez:2021qta,Borah:2021khc,Hapitas:2021ilr,Chowdhury:2021tnm,Ghorbani:2021yiw,Borah:2021jzu,Lu:2021vcp,Bai:2021bau,Das:2021zea,Kowalska:2020zve}. Also it should be noted that if we compare to the latest QCD lattice calculations this decrease to about $1.4~\sigma$\cite{Borsanyi:2020mff}. In order to clarify the origin of this difference, a new experiment
aiming to measure the contribution of hadronic corrections
is being prepared at CERN \cite{Abbiendi:2020sxw}.
One should also note that the drastic change in Hadronic Vacuum Polarization (HVP) leads to tensions elsewhere \cite{Crivellin:2020zul,Keshavarzi:2020bfy,Colangelo:2020lcg}. 

The primary purpose of direct detection experiments is the search for thermal relic particles (WIMPs) with $\rm GeV$-scale mass. However, recently, there has been significant interest in models of DM in which DM particle has a mass $m\le {\cal O}(\rm GeV)$. This is appealing because such models evade nearly all constraints on DM imposed by nucleon-DM direct detection experiments, since direct detection experiments are insensitive to the small recoil energies characteristic
of $m\le {\cal O}(\rm GeV)$.
An alternative scenario in direct search is the interaction of DM particles exclusively with the SM leptons and possibly having loop suppressed interactions with nucleons. However in this case, DM still obtain its relic abundance from Freez-out mechanism.

In this light,  direct detection strategies are proposed for DM particles with KeV to MeV mass. In
this range, DM scattering with electrons can cause single-electron
ionization signals, which are detectable with current technology. Ultraviolet photons, individual ions, and heat are interesting alternative signals. For example, the light DM regime ($1~\rm MeV$ to $10~\rm GeV$) where DM interacts with atomic electrons via recoiling is now significantly constrained by recent  data from the XENON1T\cite{XENON:2020rca}, DarkSide50\cite{DarkSide:2018ppu} and SENSEI\cite{SENSEI:2020dpa} experiments.

The goal of this paper is to broaden the scope of BSM physics. We propose light vector DM candidates with mass regime ($1~\rm MeV$ to $10~\rm GeV$). The interaction of DM with SM is leptophilic and DM particles only interact with SM leptons via spinor mediators. Scattering of vector DM from electron could be consistent with XENON1T \cite{XENON:2020rca} experiment bounds for DM-electron direct detection. The leptophilic models are particularly interesting when we look at the anomalous magnetic moment of the leptons. In this way, we attempt to find out parameter space of the model to explain the newest muon magnetic moment anomaly. 

The outline of this paper is as follows: In Section 2, we introduce framework for a light vector DM. In Section 3, we study anomlous magnetic of muon in context of the model. We consider relic density and DM-electron recoil bounds on parameters of the model in section 4 and 5. The Section 6 contains the influence of various experimental constraints on the model parameters. Finally, we present combined result in last section.

\section{The Model} \label{sec2}
Here, we present a vector DM model,
which extends the SM with massive spinor mediators which couples to vector dark matter (VDM) particles and SM leptons. The only renormalizable interaction is taken as follows:
\begin{align}\label{lagrangian}
\quad {\cal{L}} \supset \sum_{\ell=e,\mu,\tau}X_{\mu} \overline{\psi}_l \gamma^{\mu}[ g_{s}+g_{p}\gamma^{5}] \ell + {\text{H.C.}} ,
\end{align}
where $ X_{\mu} $ is the VDM candidate, and $ \psi_l $s are (Dirac) spinor mediators. 
Generally, the interaction terms $ X_{\mu} \overline{\psi}_l \gamma^{\mu}[ g_{s \psi}+g_{p \psi}\gamma^{5}] \psi_l $ and $ X_{\mu} \overline{l} \gamma^{\mu}[ g_{s l}+g_{p l}\gamma^{5}] l $ can also exist. However, these interactions will lead to $ X_{\mu} $ being unstable. To avoid such interactions we impose a $ Z_2 $ discrete symmetry under which both $ X_{\mu} $ and $ \psi_l $ are odd but all SM fields are even.
In our analysis, we have assumed universal couplings between the mediators and the SM leptons. For simplicity, we have eluded mixing between $ \psi_l $s and lepton generations.
If different lepton flavors couple with a same $\psi$ and $X_\mu$, then at one-loop level we should face the charged lepton flavor violation (CLFV) problem (for a review on the muon anomalous magnetic moment and the quest for lepton flavor violation see \cite{Lindner:2016bgg}). In such a scenario,
at one-loop level, we should have muon to electron transition process $\mu \rightarrow  e+\gamma$ which has been stringently constrained. In (\ref{lagrangian}), to avoid CLFV, we have considered different $ \psi_l $s for each lepton flavor. However, to keep the model minimal, we suppose all spinor mediators have the same mass.
In order to avoid DM decay, we suppose $M_{\psi} > M_{X}$. By considering this constraint $X_{\mu}$ will be stable and can serve as DM. Since $X_{\mu}$ is
neutral with no electric charge, for spin 1/2 mediators, the spinor $\psi_l$s have positive electric charges (and therefore couple to photon) that is equal but opposite to the charged
leptons. Therefore, $X_{\mu}$ and neutrinos cannot couple together via charged spin 1/2 mediators. VDM stability commands that no tree level mixing between the Z-boson and $ X_{\mu} $ is allowed. Therefore in our model, we have 4 independent parameters $M_{X}$, $M_{\psi}$, $g_s$ and $g_p$. Finally, we emphasize that our  approach in this paper, is quite phenomenological. We have not considered gauge invariance of Lagrangians under some gauge group, and the model is not UV-complete. For example, we supposed that $ \psi_l $s are $ SU(2)_{\text{Weak}} $ singlet while have $ U(1)_{\text{EM}} $ charge; in constructing a gauge invariant model such assumption should be considered more carefully.

\section{Anomalous Magnetic Moment (AMM) of the Muon} \label{sec5}

As it is well-known, there are some discrepancies between SM predictions and measurements in low energy experiments. One long-standing discrepancy is the anomalous magnetic dipole moment of muon:
\begin{equation}
a_\mu =(g_{\mu}-2)/2. \label{deltamagnetic1}
\end{equation}
Comparisons with experimental  measurements $a_{exp}^l$ results in studies of the magnetic moments of leptons being a powerful indirect search of new physics. The SM prediction of the anomalous magnetic moment is determined from the sum of all sectors of the
SM which include the QED contributions, the electro-weak contributions, the hadronic vacuum polarization contributions and hadronic light-by-light scattering contributions. The uncertainty of $a_{SM}$
is completely dominated by the hadronic contributions due to the non-perturbative nature of the low energy strong interaction. It is estimated from experimental measurements of the ratio of hadronic to muonic cross sections in electron-positron collisions\cite{Davier:2010nc,Aoyama:2012wk,Aoyama:2019ryr,Czarnecki:2002nt,Gnendiger:2013pva,Davier:2017zfy,Keshavarzi:2018mgv,Colangelo:2018mtw,Hoferichter:2019mqg,Davier:2019can,Keshavarzi:2019abf,Kurz:2014wya,Melnikov:2003xd,Masjuan:2017tvw,Colangelo:2017fiz,Hoferichter:2018kwz,Gerardin:2019vio,Bijnens:2019ghy,Colangelo:2019uex,Blum:2019ugy,Colangelo:2014qya,Aoyama:2020ynm}. The combined results of muon anomalous magnetic moment in two experiments at FermiLab (FNAL E989)\cite{Muong-2:2021ojo}, and Brookhaven National Laboratory (LabBNL E821) \cite{Muong-2:2006rrc} yield overall deviation from the SM central value: 
\begin{equation}
\Delta a_\mu =a^{exp}_\mu-a^{\rm SM}_\mu =(25.1\pm 5.9)\times 10^{-10}, \label{deltamagnetic}
\end{equation}
which corresponding to a significance of $4.2~\sigma$. 

In light of the results, we study constraints on the parameter space of the model. Since in our model, VDM particles interact with the SM leptons via a massive mediator, a significant effect on the anomalous magnetic moment of the leptons is expected.
 
\begin{figure}%[!htb]
	\begin{center}
		\centerline{\hspace{0cm}\epsfig{figure=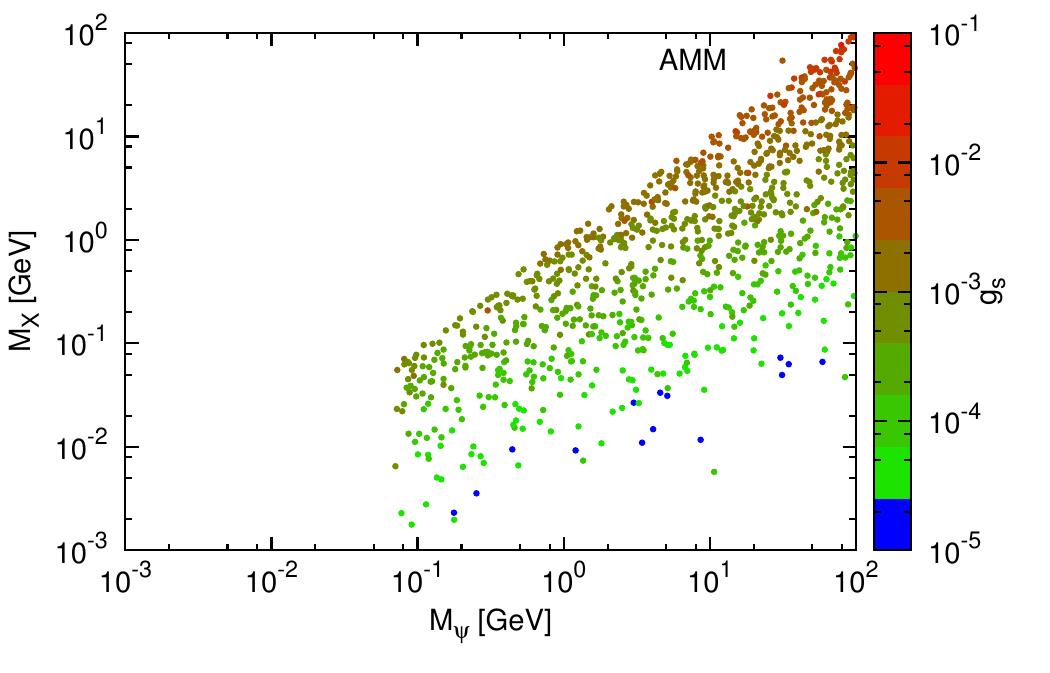,width=8cm}\hspace{0.5cm}\hspace{0cm}\epsfig{figure=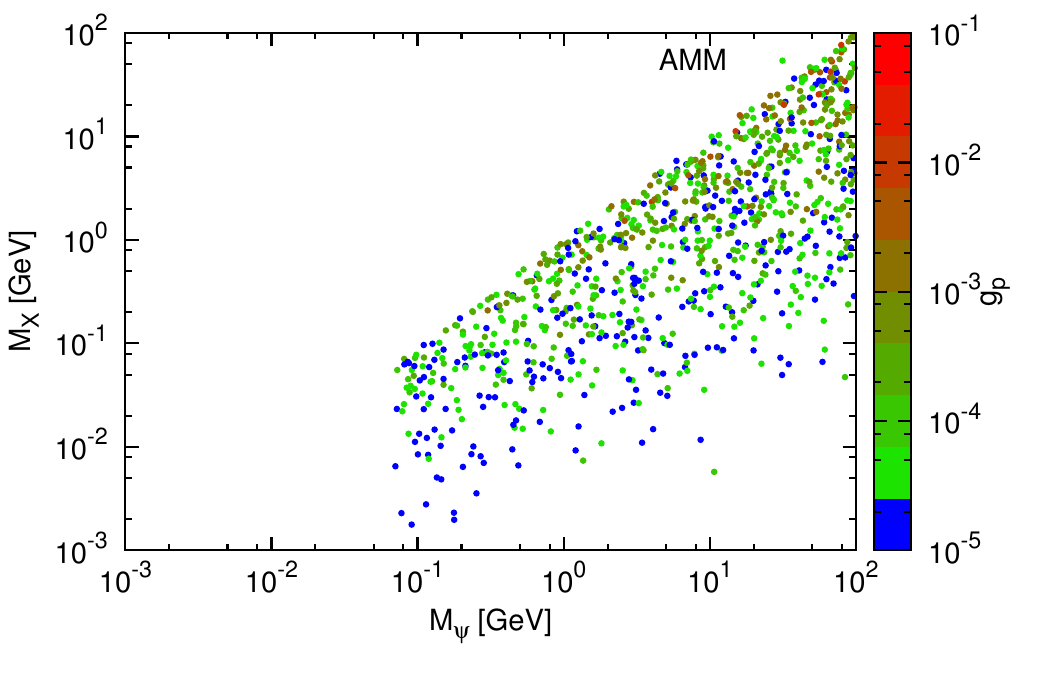,width=8cm}}
		%\centerline{\vspace{0.2cm}\hspace{-0.7cm}(a)\hspace{8cm}(b)}
		\centerline{\vspace{-0.2cm}}
		\caption{Scatter-points consistent with AMM of the Muon. We scanned over all free parameters of the model.
Note that, the parameter space could be limited by other experiments, such as BaBar, Belle II and LEP II. However, maybe the case of a small mass difference of $ M_\psi – M_X = \delta > 0 $ with $ \delta / M_\psi \ll 1 $ is still allowed by experiments (with large missing energy plus soft electrons/muons), like the case of supersymmetry. In addition, if $ 2 M_\psi $ is lighter than the tau lepton $ m_\tau $, the decay spectrum and width of tau lepton will be changed. Thus, the parameter space left are smaller than that in this figure. To see how these experiments limit the parameter space see, e.g., \cite{BaBar:2016sci,BaBar:2020jma,Belle-II:2019qfb}} \label{amu}
	\end{center}
\end{figure}

In the present work, as we mentioned before, we consider universality for all interactions of leptons. Therefore, we only consider the magnetic moment of the muon and ignore weaker constraints on the  AMMs of tau and electron.
The model contribution to muon magnetic moment arises from the one-loop  mediated by VDMs. The analytical expression for $\Delta a_{\mu}^{\psi}$ is \cite{Leveille:1977rc}-\cite{Queiroz:2014zfa}:
\begin{equation}
\Delta a_{\mu}^{\psi}=\frac{1}{8\pi^2} \frac{m_{\mu}^2}{M_{X}^2}\int\frac{g_s^2P_s(x)+g_p^2P_p(x)}{(1-x)(1-\lambda^2 x)+\epsilon^2\lambda^2x}dx
\end{equation}\label{expersion}
where 
\begin{align}
P_s(x)&=2x(1-x)(x-2(1-\epsilon)^2)+\lambda^2(1-\epsilon)^2x^2(1+\epsilon-x)\nonumber\\
P_p(x)&=2x^2(1+x+2\epsilon)+\lambda^2(1-\epsilon)^2x(1-x)(x-\epsilon)\nonumber
\end{align}\label{cofficients}
and $\epsilon=M_{\psi}/m_{\mu}$, $\lambda=m_{\mu}/M_X$ ($m_{\mu}$ is the muon mass, $M_{\psi}$ and $M_{X}$ are spinor mediator and DM mass). Therefore the contribution of a generic spinor mediator to the muon anomalous magnetic moment is given by:
\begin{align}
\Delta a_{\mu}^{\psi} =\left(\frac{m_{\mu}} {2\pi M_{X}}\right)^2\left\lbrace\left[\frac{ M_{\psi} }{m_{\mu}} -\frac{2}{3}\right]g_{s}^2+\left[-\frac{ M_{\psi}}{m_{\mu}} -\frac{2}{3}\right] g_{p}^2\right\rbrace, \label{formula magnetic}
\end{align}
where $g_s$ and $g_p$ are couplings of the SM leptons with new fields in accordance with interaction terms of Eq.~\eqref{lagrangian}. 
Figs.~\ref{amu}-a and b show allowed region in $M_X-M_{\psi}$ plane for random values of couplings which is consistent with BNL E821 and FNAL E989 experiments. As it is seen in figures, for masses of spinor and DM between a few hundred $\rm MeV$ to $100~\rm GeV$, the contribution of the model can explain $a_{\mu}$ anomaly.  Note that in each plot, we turn on both couplings ($g_s$ and $g_p$).

Before the next section, we comment on the electron g-2 result as a comparison.
There is a 2.4 $ \sigma $ discrepancy between the theoretical prediction \cite{Aoyama:2017uqe} and the existing experimental measurement \cite{Hanneke:2008tm,Hanneke:2010au} of the electron anomalous magnetic moment,
\begin{equation}
\Delta a_e =a^{exp}_e-a^{\rm SM}_e =(-87\pm 36)\times 10^{-14}. \label{elect}
\end{equation}
Given the negative sign of the above relation, to explain both electron and muon g-2 in this model, according to (\ref{formula magnetic}), one should consider different couplings for electron and muon generations. Otherwise, $ \Delta a_{\mu}^{\psi} > 0 $ leads to $ \Delta a_{e}^{\psi} > 0 $.
However, considering the significance of 4.2 $ \sigma $ of muon anomalous magnetic moment and ignoring 2.4 $ \sigma $ electron g-2 anomaly, we continue to assume universal couplings between the mediators and the SM leptons.

\section{Relic Density of DM}

Cosmological observations point toward the existence of DM, whose cosmological relic abundance is approximately a fourth of the energy budget of the Universe. According to the recent
 measurements from the PLANCK \cite{Planck:2018vyg} collaboration DM relic density is:
\begin{equation}
\Omega_{DM} h^2 = 0.120 \pm 0.001 ,
\end{equation}
where $ h\sim0.7 $ is the Hubble expansion rate at present times in units of 100 $ (km/s)/Mpc $.

\begin{figure}%[!htb]
	\begin{center}
		\centerline{\hspace{0cm}\epsfig{figure=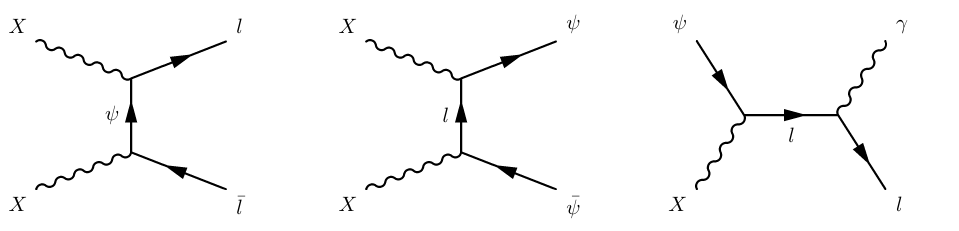,width=10cm}}
	\centerline{\vspace{0.5cm}\hspace{0cm}(a)\hspace{2.8cm}(b)\hspace{2.8cm}(c)}
		\centerline{\vspace{-0.7cm}}
		\caption{Feynman diagrams for DM annihilation into (a) leptons and (b) spinor mediators. For every t channel there is associated u channel not depicted here. DM coannihilaion is depicted in (c).}\label{FD}
	\end{center}
\end{figure}

We suppose the mechanism that generates DM
cosmological relic density is thermal freeze-out described by a Boltzmann equation of the form
\begin{equation}
\frac{d n_{\chi}}{dt} + 3H n_{\chi}= - \langle\sigma_{eff} |v_{rel}|\rangle (n_{\chi}^{2} - n_{\chi,eq}^{2}),
\end{equation}
where $ n_{\chi} $ is the number density of vector DM, $ H $ is the Hubble parameter, and $ \langle\sigma_{eff} |v_{rel}|\rangle $ is the thermally averaged of effective cross section including both DM annihilation and coannihilaion channels \cite{Hooper:2009zm} (see figure \ref{FD}).

\begin{figure}%[!htb]
	\begin{center}
		\centerline{\hspace{0cm}\epsfig{figure=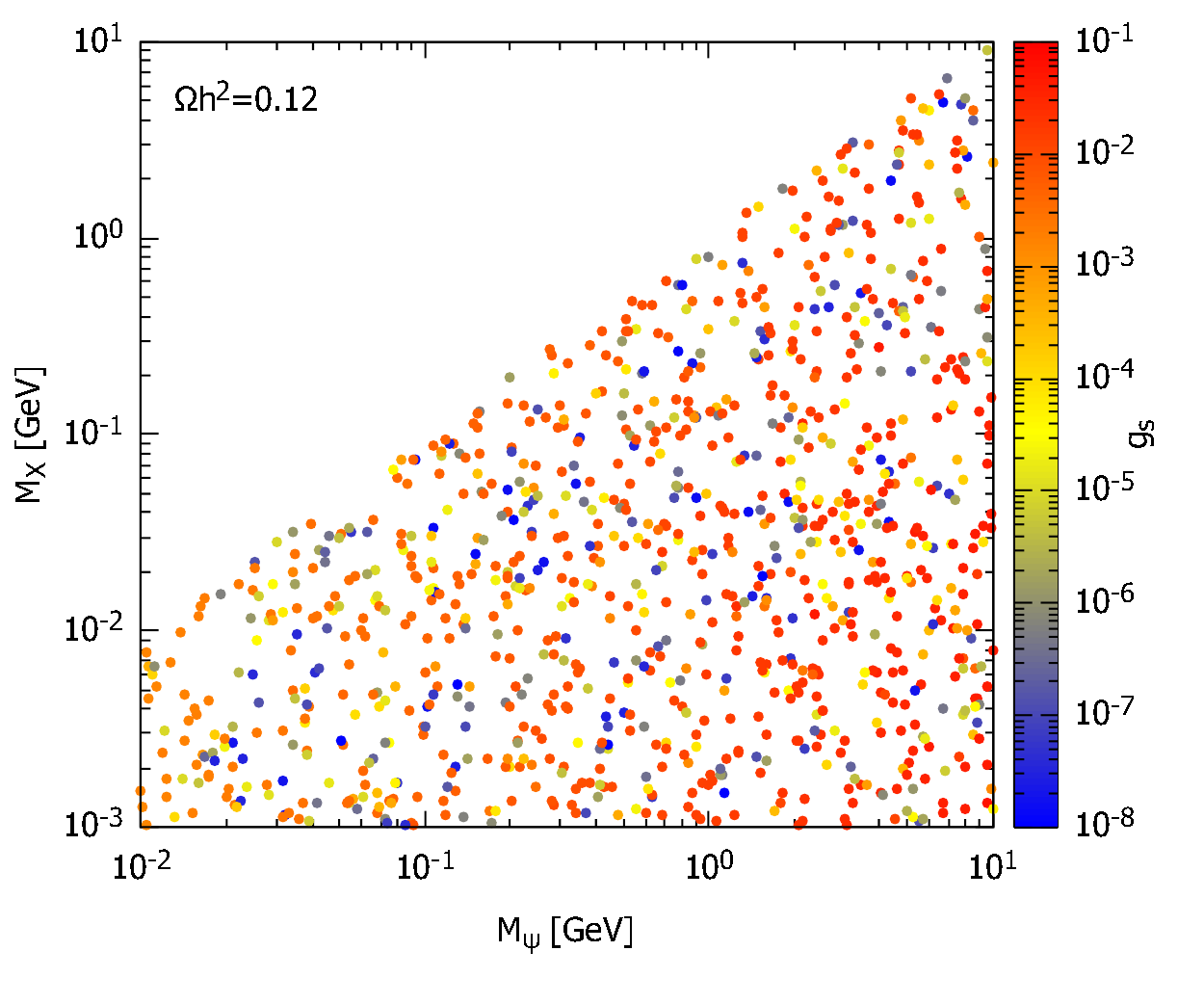,width=8cm}\hspace{0.5cm}\hspace{0cm}\epsfig{figure=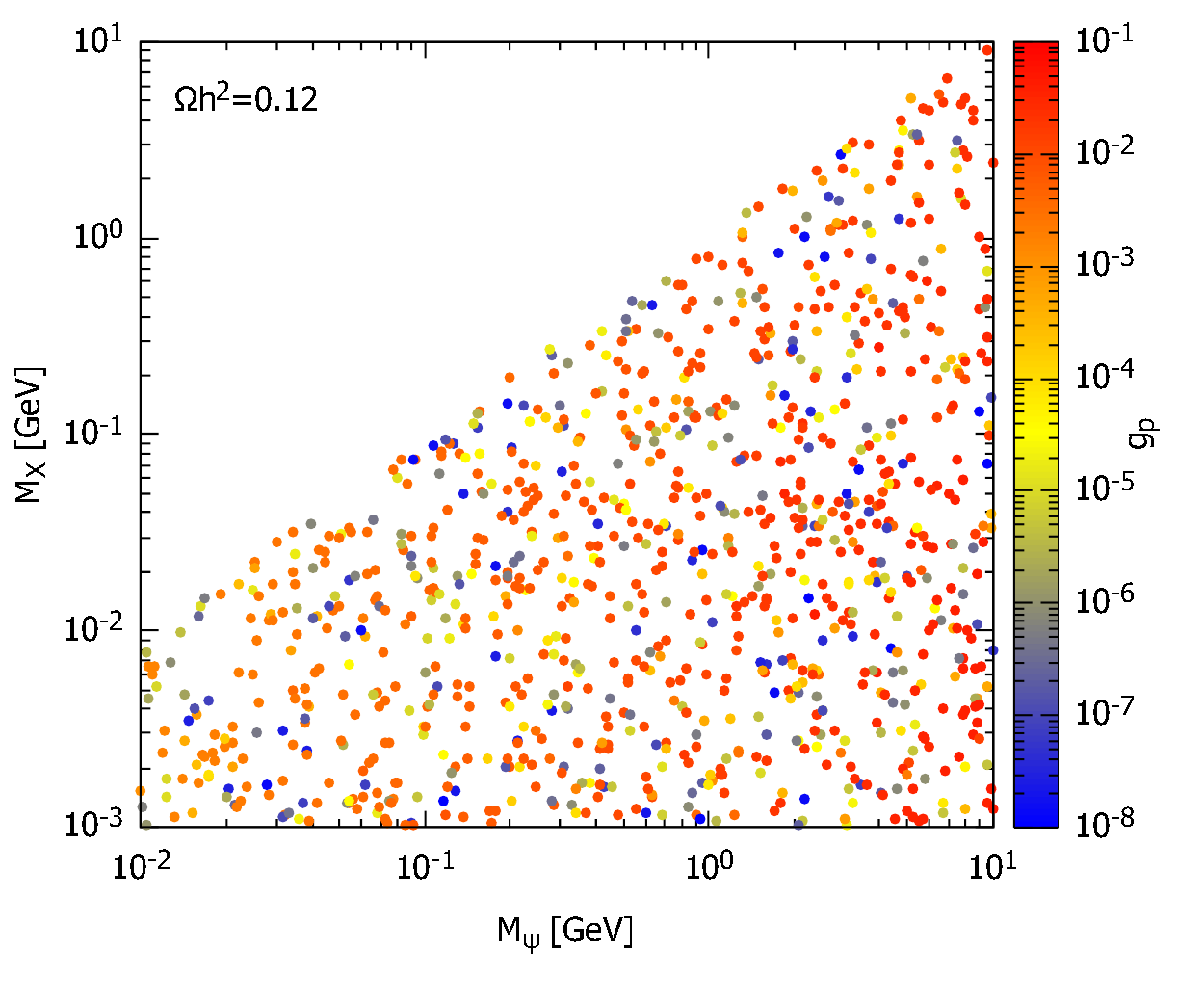,width=8cm}}
		%\centerline{\vspace{0.2cm}\hspace{-0.7cm}(a)\hspace{8cm}(b)}
		\centerline{\vspace{-0.2cm}}
		\caption{Dark matter relic density consistent with Planck data. We scanned over all free parameters of the model}\label{Relic}
	\end{center}
\end{figure}

The constraint from  DM relic density on the vector DM have
been determined by implementing the model in the numerical package {\tt MicrOMEGAs} \cite{Barducci:2016pcb}. The allowed parameter space corresponding to this constraint is depicted in figure \ref{Relic}.

Note that the annihilation channel in Fig. 2 (b) is phase-space suppressed and the main channel is Fig. 2 (a). The s-wave annihilation $ X \, X\rightarrow e^+  \, e^- $ is restricted by the CMB observation (see, e.g. \cite{Slatyer:2015jla}) for the mass of dark matter about $ < 10 $ GeV, and the AMS-02 observation (see, e.g. \cite{Bergstrom:2013jra,Liu:2014cma}) for the mass of dark matter from a few GeV to 100 GeV. There is nearly no parameter space left with joint constraints. Thus, the relic density of $ X $ equal to 0.12 seems to be in tension with CMB and AMS-02 observations. To resolve this issue, we consider coannihilation effect with $ M_{\psi} - M_{X} = \delta > 0 $ and $ \frac{\delta}{M_{\psi}} \ll 1 $.
The particles $ X $ and $ {\psi}_l $ can be in thermal equilibrium for a while in the early universe via processes like Fig. 2 (b), and the abundance of $ X $ will be reduced by processes such as Fig. 2 (c).
However, in our combined analysis, Sec. 7, we see that the relic density of $ X $ should be smaller than 0.12. Therefore, we consider $ \Omega_{X} h^2 < 0.12 $ as DM relic density constraint. This means $ X $ cannot be the only DM component.

\section{Direct Detection and Electron recoil}
\begin{figure}[!htb]
	\begin{center}
		\centerline{\hspace{0cm}\epsfig{figure=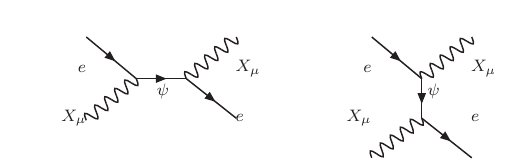,width=10cm}}
		\caption{DM-electron Feynman diagram}\label{Direct Detection electron}
	\end{center}
\end{figure}

Let us now turn our attention to the direct detection possibilities of DM in the framework.  As it is mentioned, we consider the hypothesis that the VDM particle $X_{\mu}$ only couples directly to leptons in particular the electrons but not to quarks. For this reason, DM-nucleon interaction is absent at tree level. However, DM-electron elastic scattering of spin-independent type is feasible at tree level. In the following, we ignore the loop suppressed DM-nucleon interaction. In comparison to underground experiments focused on neutrino detection, that are typically sensitive to energy depositions above a few hundred $\rm keV$, Xenon-based dark matter detectors provide the leading sensitivity for electronic energy depositions of $100~\rm keV$ and below.  In our model the relevant interaction is described by the Lagrangian \ref{lagrangian}. Electron recoil can occur corresponding to Feynman diagrams in Fig.~\ref{Direct Detection electron}. Therefore, in non-relativistic limit the elastic scattering cross section of the DM-electron has following form\cite{Essig:2011nj}:
\begin{align}
\sigma_{DM-e}\approx\frac{g_{s}^4\mu_{eX}^2}{2\pi M_{X}^2M_{\psi}^2} ,  \label{Direct Detection electron1}
\end{align}
where $\mu_{eX}$ is the DM-electron reduced mass. The cross section that includes $g_p$ is zero due to the odd number of $\gamma^ 5$ in the trace. 
So far, in direct detection experiments there is found no evidence of DM-electron elastic scattering. Nevertheless several experiments have set upper bounds on this cross section:  XENON1T\cite{XENON:2020rca}, DarkSide50\cite{DarkSide:2018ppu} and SENSEI\cite{SENSEI:2020dpa}. We consider upper bound from the XENON1T experiment to search for DM interacting with electrons\cite{XENON:2020rca}.

\section{Additional constraints on the leptophilic VDM interaction }
Thus far in this paper, we have included discussion
	of the relic density, direct detection and $(g-2)_{\mu}$   constraints. In this section, we look at the various experimental constraints on the muon-VDM interaction in the model. Two kinds of these experimental constraints are as follows:
\begin{itemize}

\item Big Bang Nucleosynthesis (BBN) constrains the DM physics by three generic ways, the number of the excited relativistic degrees of freedom, energy injection due to annihilation or decay of heavy particles, and particle catalysis of nuclear reactions. For the first case, traditionally, this procedure is interpreted in terms of a bound on the effective number of neutrino. Note that in our model, $X_{\mu}$ is neutral and  cannot couple neutrino via charged spin 1/2 mediators.  In a more general way, this bound can be translated into a limit on dark radiation. In \cite{Pospelov:2010hj}, it have been shown that BBN constraint as a limit on dark radiation which is massless or nearly-massless degrees of freedom that have the expected scaling as the Universe expands $\rho_{dr}\approx~a^{-4}$.  They assume that the  additional relativistic component does not exchange energy with any SM species. This viewpoint can then be easily applied to the thermally decoupled extra degrees of freedom.  For this reason, in our model in which new particles are not massless, the parameter space is not sensitive to the number of the excited relativistic degrees of freedom. Nevertheless, constraints on $\langle\sigma v\rangle$ from electromagnetic energy injection are proportional to temperature and dark matter mass. In \cite{Henning:2012rm}, it was shown that BBN procedure rules out DM annihilation to $e^+e^-$ for the DM mass region  $30~\rm MeV \lesssim m_{\chi}\lesssim 77 ~\rm MeV$.

\item The observation of cosmic charge leptons have reported in wide range of the data ($10~\rm GeV$-$10~\rm TeV$); Fermi-LAT has reported a measurement of the CR electron-positron spectrum from $7~\rm GeV$ to $2~\rm TeV$ \cite{Fermi-LAT:2017bpc}. The PAMELA satellite experiment\cite{PAMELA:2008gwm} released an abundance of the positron in the CR energy range of $15-100~\rm GeV$.  Also the results of a CR electron-positron spectrum, between $10~\rm GeV$ and $\rm 3 TeV$, have been reported based upon observations with CALET \cite{CALET:2017uxd}. The DAMPE collaboration \cite{DAMPE:2017cev,DAMPE:2017fbg} has been presented a measurements of the electron-positron spectrum in the energy range $25~\rm  GeV$ to $4.6 ~\rm  TeV$. These observations are both considered as signs of standard astrophysical sources or DM annihilation. In our model, DM has mass smaller than $10~\rm GeV$  then the parameter space is not sensitive to these observations.
\end{itemize}

\section{Combined Results}
\begin{figure}%[!htb]
	\begin{center}
		\centerline{\hspace{0cm}\epsfig{figure=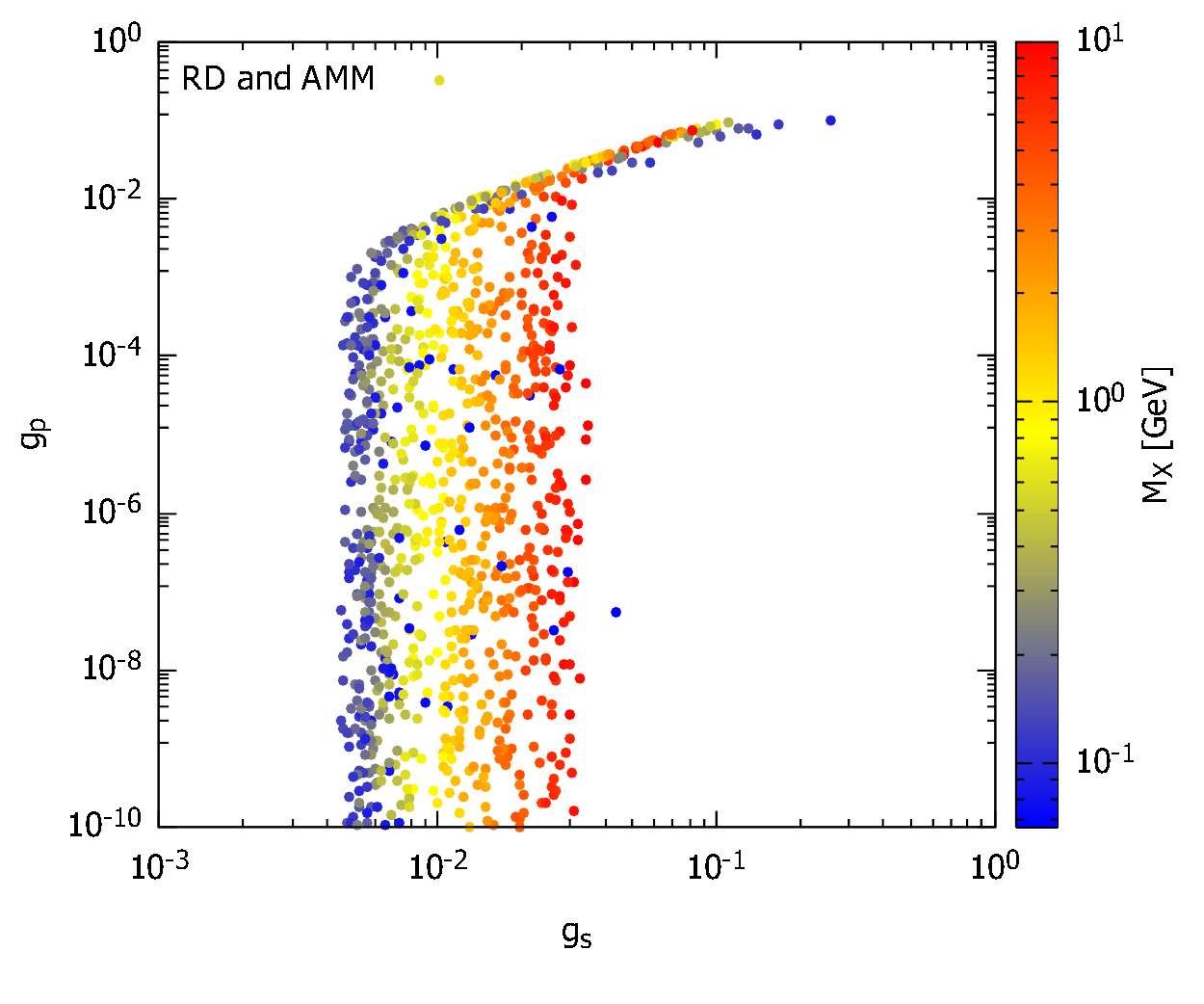,width=8cm}\hspace{0.5cm}\hspace{0cm}\epsfig{figure=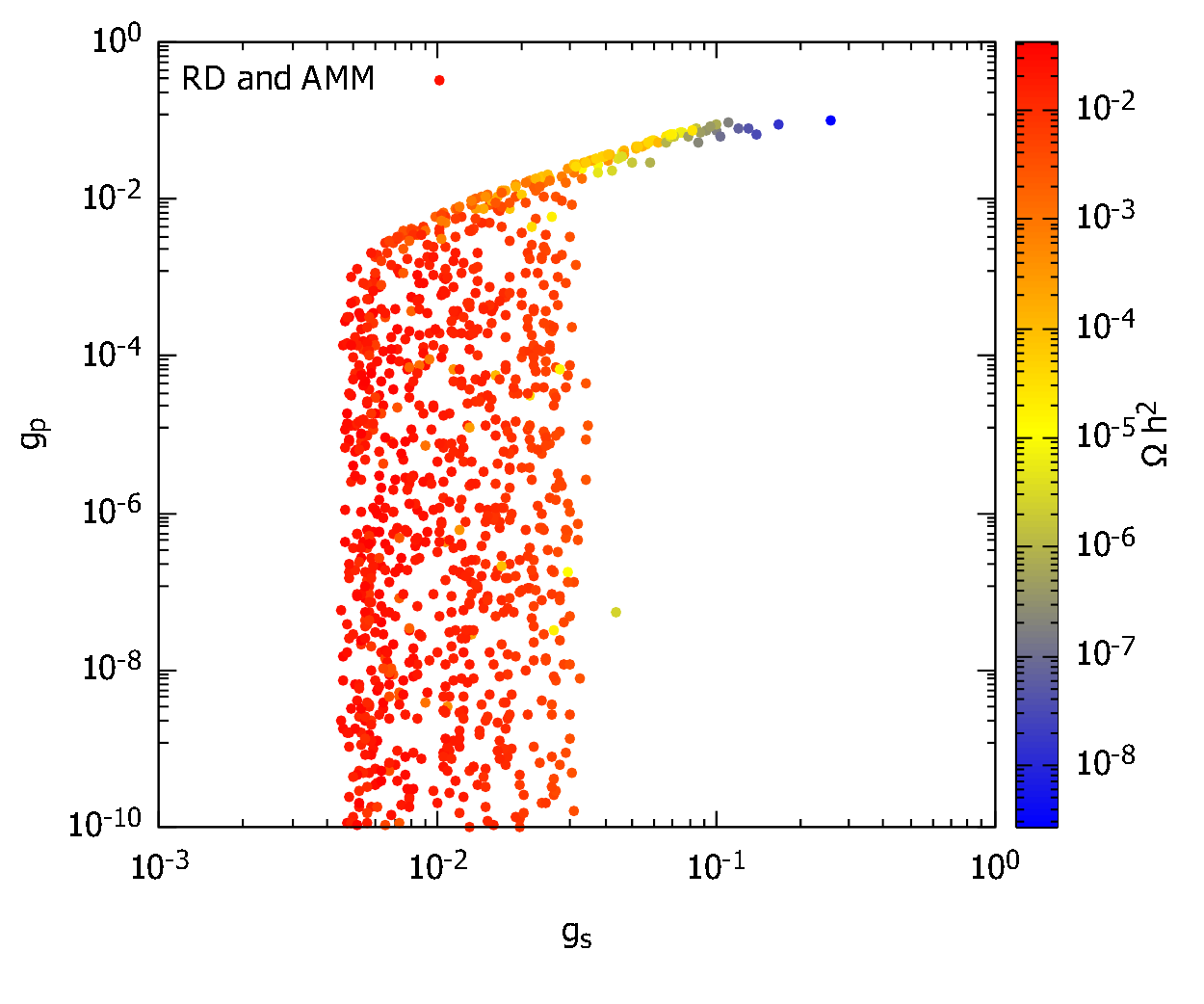,width=8cm}}
		\centerline{\vspace{0.2cm}\hspace{-0.7cm}(a)\hspace{8cm}(b)}
		\centerline{\vspace{-0.2cm}}
		\caption{Scatter-points consistent with Anomalous Magnetic Moment (AMM) of the Muon and DM relic density (RD).}\label{amu1}
	\end{center}
\end{figure}

In this section, we present a combined analysis of all observational constraints described in the previous sections. The constraints discussed are summarized by Fig.~\ref{amu1}. As it is seen, for narrow region in parameter space, all constraints are satisfied. Remarkable point is for light DM (${\cal O}~(\rm MeV)$), implying we can have thermal DM which explain anomalous magnetic moment of the muon.  

\begin{figure}%[!htb]
	\begin{center}
		\centerline{\epsfig{figure=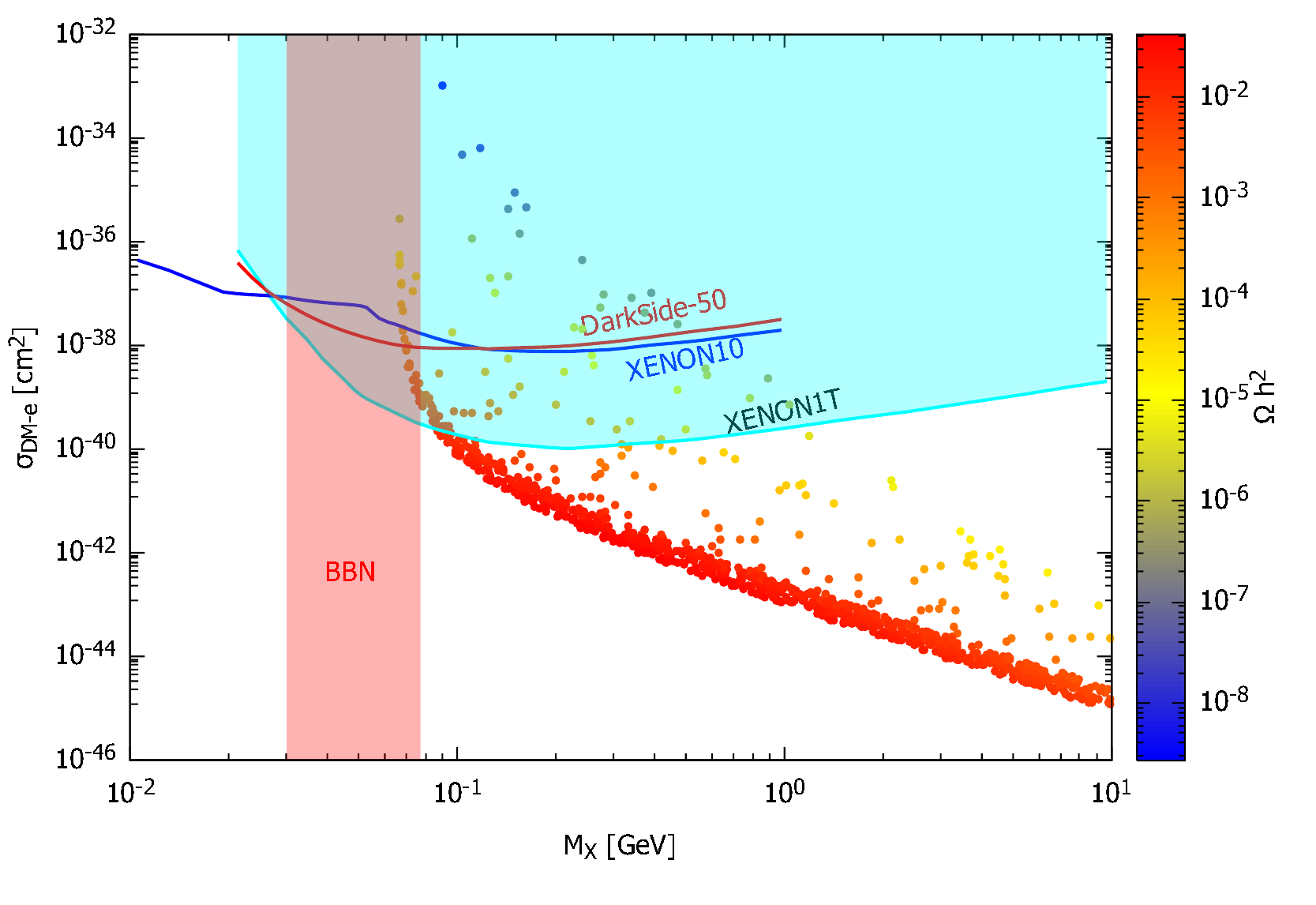,width=12cm}}
		\caption{Scatter-points consistent with AMM of the Muon and DM relic density (RD). Some points will be excluded by XENON1T upper bound on DM-electron cross section. Points that are located in a narrow red area are excluded by BBN.}\label{combined}
	\end{center}
\end{figure}

To obtain the parameter space consistent with all constraints, i.e., muon anomalous magnetic moment, DM relic density, and DM-electron direct detection constraints, first we solve Eq.~(\ref{formula magnetic}) with respect to $ g_s $:
\begin{equation}
g_s = \sqrt{\frac{12 \pi^2 M_X^2}{m_{\mu}(3 M_{\psi} - 2 m_{\mu})} \Delta a_{\mu }^{\psi} + \frac{3 M_{\psi} + 2 m_{\mu}}{3 M_{\psi} - 2 m_{\mu}}g_p^2}
\label{psi mass}
\end{equation}
Since $ \text{$\Delta $a}_{\mu }^{\psi } $ is positive, $ M_{\psi } > \frac{2}{3} m_{\mu } $, and therefore $ g_s > g_p $. Now, to obtain the parameter space consistent with DM relic density ($ \Omega_{X} h^2 < 0.12 $), we scan over random points for $ M_{\psi} $, $ g_p $ with the extra condition  $M_{\psi} - m_e - \Delta < M_{X} < M_{\psi} - m_e$ (where $ M_{X} $ is chosen randomly and we put $ \Delta = 10^{-6} $) and for $ g_s $ we use Eq.~(\ref{psi mass}) where $ \text{$\Delta $a}_{\mu }^{\psi } $ is chosen randomly in a domain given by Eq.~(\ref{deltamagnetic}). In this way, we obtain the parameter space which gives the correct value of both DM relic density and muon anomalous magnetic momont. Finally we keep the points that are also consistent with direct detection constraint. The result of our combined analysis is depicted in Fig.~\ref{combined}.

\section{Conclusions} \label{sec7}
SM has achieved great success for its high accuracy to
describe electroweak and strong interaction. However,
there remains problems such as DM which SM can not explain. In addition,
recent results about muon anomalous magnetic moment
 brings new challenges to the SM. The $4.2~\sigma$
discrepancy between experiment and SM prediction
seems to indicate new physics behind ($g_{\mu}-2$) anomaly
and gives possible hints to the BSM physics. In light of these results, we consider a vector DM model with a leptophilic spinor mediator coupled to the SM charged leptons. From phenomenological
point of view, we probed the parameter space of the model for light DM ${\cal O}~(\rm MeV$) and sub-GeV spinor mediator. To conclude, the anomalies associated to the muon
physics can be successfully addressed in light VDM regime. If the anomalies are confirmed by forthcoming experimental
analyses, our results show that the model can be surveyed by DM-electron direct detection.

\providecommand{\href}[2]{#2}\begingroup\raggedright\endgroup

\end{document}